\documentclass{emulateapj}
\slugcomment{{\sc Accepted to ApJ:} October, 2012} 
\usepackage{latexsym}
\usepackage{amsmath}
\usepackage{amsfonts}
\usepackage{amssymb}
\usepackage{makeidx}
\usepackage{graphicx}
\usepackage{natbib}

\newcommand{\um}{\mbox{\,$\mu$m}}

\begin{document}
\title{Filamentary Star Formation: Observing the Evolution toward Flattened Envelopes}
\author{Katherine Lee, Leslie Looney}
\affil{Department of Astronomy, University of Illinois at Urbana-Champaign, 1002 W Green St, Urbana, IL 61801, USA; ijlee9@illinois.edu, lwl@illinois.edu}
\author{Doug Johnstone}
\affil{Department of Physics and Astronomy, University of Victoria, P.O. Box 3055, STN CSC, Victoria, BC V8W 3P6, Canada; Douglas.Johnstone@nrc-cnrc.gc.ca}
\affil{NRC-Herzberg Institute of Astrophysics, 5071 West Saanich Road, Victoria, BC V9E 2E7, Canada}
\author{John Tobin}
\affil{Hubble Fellow, National Radio Astronomy Observatory, Charlottesville, VA 22903, USA; jtobin@nrao.edu}

\begin{abstract}
Filamentary structures are ubiquitous from large-scale molecular clouds (few parsecs) to small-scale circumstellar envelopes around Class 0 sources ($\sim$1000 AU to $\sim$0.1 pc). 
In particular, recent observations with the \textit{Herschel Space Observatory} emphasize the importance of large-scale filaments (few parsecs) and star formation.  
The small-scale flattened envelopes around Class 0 sources are reminiscent of the large-scale filaments.
We propose an observationally derived scenario for filamentary star formation that describes the evolution of filaments as part of the process for formation of cores and circumstellar envelopes. 
If such a scenario is correct, small-scale filamentary structures (0.1 pc in length) with higher densities embedded in starless cores should exist, although to date almost all the interferometers have failed to observe such structures.  
We perform synthetic observations of filaments at the prestellar stage by modeling the known Class 0 flattened envelope in L1157 using both the Combined Array for Research in Millimeter-wave Astronomy (CARMA) and the Atacama Large Millimeter/Submillimeter Array (ALMA).  
We show that with reasonable estimates for the column density through the flattened envelope, the CARMA D-array at 3mm wavelengths is not able to detect such filamentary structure, so previous studies would not have detected them. 
%We further model starless cores as density profiles consistent with Bonner-Ebert profiles, and suggest that under the constraint of a few solar masses for the total mass, structures larger than 900 AU (in radius) for spherical models and 450 AU (in length) for filamentary models are not detectable with the CARMA D-array at 3mm continuum.  
However, the substructures may be detected with CARMA D$+$E array at 3 mm and CARMA E array at 1 mm as a result of more appropriate resolution and sensitivity.
ALMA is also capable of detecting the substructures and showing the structures in detail compared to the CARMA results with its unprecedented sensitivity.  
Such detection will confirm the new proposed paradigm of non-spherical star formation. 

\end{abstract}

\section{Introduction}

It is becoming clear that filamentary structures (few parsecs to 10 parsecs in length and typically 0.1 pc in width)
in molecular clouds are common and need to be understood.
One clear example is the Integral-Shaped Filament region in the north of Orion-A, comprising OMC 1-4, where prestellar cores and protostars are forming \citep[e.g.,][]{1997ApJ...474L.135C,1999ApJ...510L..49J, 2000ApJS..131..465A,2007MNRAS.374.1413N,2007ApJ...665.1194I,2008ApJ...688..344T}.
Taurus also consists of several large filaments, each with ongoing star-forming activity \citep[e.g.,][]{1995ApJ...445L.161M,1998ApJ...502..296O, 2008hsf1.book..405K}.
The large-scale filaments are even more evident in recent observations from the \textit{Herschel Space Observatory} \citep[e.g.,][]{2010A&A...518L.102A,2010A&A...518L.103M,2011A&A...529L...6A,2011A&A...533A..94H}, and these observations further suggest a tight connection between the formation of dense cores and gravitationally unstable filaments.
%In particular, recent observations with the \textit{Herschel Space Observatory} further suggest a tight connection between the formation of dense cores and gravitationally unstable filaments \citep[e.g.,][]{2010A&A...518L.102A,2010A&A...518L.103M,2011A&A...529L...6A,2011A&A...533A..94H}.
While the mechanisms for forming these filamentary structures are still under debate \citep[e.g.,][]{2004RvMP...76..125M,2008ApJ...674..316H,2008ApJ...687..354N,2009ApJ...700.1609M,2011ApJ...735...82M,2011ApJ...740...88P}, an observationally derived process has been suggested:
first, the filamentary structures at large scales form, possibly as a result of magnetic-hydrodynamic (MHD) turbulence in the ISM, and secondly, the prestellar cores form from the fragments of a subset of filaments through gravitational instability.

As the role of large-scale filaments in molecular clouds has received significant attention, a number of the latest observations have unveiled filamentary structure at smaller scales ($\sim$ few tenths of parsecs in length and 100ths of parsecs in width).  
For instance, \citet{2011A&A...533A..34H} have observed four subsonic, velocity-coherent filaments in L1517 ($\simeq$0.5 pc in length) that are possibly condensed out from the more turbulent natal cloud and lead to the quasi-static fragmentation of cores.  
In addition, \citet{2011ApJ...739L...2P} probed the Barnard-5 star-forming core with high angular resolution and discovered filamentary structures with $\sim$ 0.1 parsecs in length.  
The filaments in Barnard-5 are possibly the result of fragmentation in a coherent region where subsonic motions dominate, and are likely to form stars via future gravitational collapse.  

In addition to the filamentary structures in molecular clouds at the early stage of star formation, typical length of few parsecs and width of 0.1 pc, small-scale filamentary structures, typical length of few thousand AU to 0.1 pc and width of few hundred to few thousand AU, have also been observed in the envelopes around Class 0 protostars \citep{2010ApJ...712.1010T}.  
%In particular, \citet{2007ApJ...670L.131L} revealed the first clear detection of a flattened circumstellar envelope around the young Class 0 source in L1157 in Cepheus. 
These filamentary structures in the protostellar envelopes are mostly irregular and non-axisymmetric in morphology, 
suggesting the initial non-equilibrium from the prestellar stage. 
The filamentary structure presented near the Class 0 source is reminiscent of the large \textit{Herschel} observed structures, 
although the size scales of the two are distinct and the properties are presumably different.  

The relationship between the large-scale filaments in molecular clouds and small filamentary envelopes around young protostars still requires further investigation.
Several numerical simulations have shown that large-scale filaments in molecular clouds are prone to fragmentation leading to prestellar cores \citep[e.g.,][]{1997ApJ...480..681I,2002ApJ...578..914H}, and filaments are possibly the most favorable mode for fragmentation \citep{2011ApJ...740...88P,2012ApJ...756..145P}.  
Moreover, studies have also demonstrated that filamentary geometries at large scales have a significant impact on the geometries and symmetries of the subsequently collapsing cores \citep{2011MNRAS.411.1354S}.  

These observations and numerical simulations deliver a clear message: 
filamentary structures from large to small scales are clearly playing an important role to the star formation process.
In this paper, we suggest an observational evolution between filaments at the large scale and
filaments on the small scale. 
The proposed scenario suggests that the small-scale filamentary structure (few thousand AU) in protostellar envelopes originate from the filamentary structure (0.1 pc) embedded in the larger envelopes of starless cores instead of being produced by the protostellar collapse.
We will further show why the filamentary structures in starless cores have not been observed to date.

\section{Evolution of Protostellar Structure}

It has been well known that dust emission maps of Class 0 sources show
very spherical emission \citep[e.g.,][]{2000ApJ...529..477L,yancy2000,motte2001}.  Although
molecular surveys of dense cores showed non-spherical structures 
\citep[e.g.,][]{1991ApJ...376..561M}, these non-symmetric structures were often
considered to be material not directly involved in the star formation process, 
i.e.\ part of the larger-scale molecular cloud or clump,
so these components were rarely used in the observational modeling of these sources.
Instead many authors assumed that the spherical dust emission indicated
spherical collapse \citep[e.g.,][]{1977ApJ...214..488S,1984ApJ...286..529T}
and used this symmetry to 
derive envelope properties and 
place constraints on any embedded disk components 
\citep[e.g.][]{keene1990,looney2003,harvey2003,jess2009}.
%However, recent results have shown that mode of thinking to be incorrect.
However, recent studies have shown the envelope structures to be more complex.

\subsection{Changing the Paradigm for the Inner Envelope of Class 0 Protostars}

\begin{figure}
\begin{center}
\includegraphics[scale=1.7]{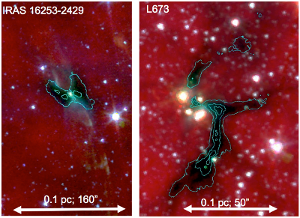}
\caption{Example of the envelopes detected with Spitzer 8 micron imaging from \citet{2010ApJ...712.1010T}.
The images are from \textit{Spitzer} 8.0 $\um$ observations.
%The 8.0 $\um$ optical depth contour levels are 0.3, 0.53 for IRAS 16253-2429 (left panel) and 0.95 0.3, 0.6, 1.2 for L673 (right panel).
}
\label{fig:rogues}
\end{center}
\end{figure}

The ability to use 8 $\um$ absorption against PAH emission background allows
the decoupling of the dust density and temperature for the first time in Class 0 sources
\citep[e.g.,][]{2007ApJ...670L.131L,2010ApJ...712.1010T}.
With these measurements, it was realized that the dense portions of the envelope
are complex, filamentary, and often non-axisymmetric 
structures ($\sim$ 1000 AU to 0.1 pc).
Figure \ref{fig:rogues} illustrates the diversity of structures seen in the Tobin sample.
IRAS 16253-2429 is what one would expect to see in a spherical envelope case. The 8 
$\um$ absorption is not a good tracer at the central source or in the outflow cavity, since 
in both cases there is emission in addition to the background.
In stark contrast, L673 is a clear example of the main point of \cite{2010ApJ...712.1010T},
which is that flattened, filamentary, and non-axisymmetric envelopes are the typical envelope structure. 

\begin{figure}
\begin{center}
\includegraphics[scale=1.2]{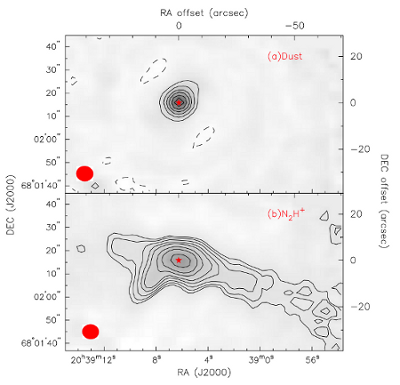}
\caption{Example of dust and N2H+ differences from \cite{2010ApJ...709..470C}.  However,
both emission was fit in a simple density model.
}
\label{fig:Chiang}
\end{center}
\end{figure}

How does this result reconcile with interferometric dust emission observations which
show spherical emission in these sources
\citep[e.g.,][]{2000ApJ...529..477L}?
It is important to remember that dust emission depends on both dust density and temperature.
With flattened or non-axisymmetric envelopes and/or outflow cavities in young sources, 
the heating will be
inhomogeneous; the lower density material near the central source
is heated more, leading to temperature and density gradients, 
and the dust emission will appear
more spherical even if the dust distribution is not.
A good example, shown in Figure \ref{fig:Chiang} from \cite{2010ApJ...709..470C}, 
is the source L1157.
Although there is a flattened and filamentary envelope detected in both N$_2$H$^+$ 
and the 8 $\um$ absorption (also seen in Figure \ref{fig:spitzer}),
the dust emission is very spherical and typical of a Class 0 protostar.
\cite{2012ApJ...756..168C} constructed a model that has a flattened geometry similar to the 
N$_2$H$^+$ and 8 $\um$ absorption features and yet still predicts the observed 
spherical dust continuum when non-spherical, self-consistent temperature solutions are used.

\begin{figure}
\begin{center}
\includegraphics[scale=0.38,angle=270]{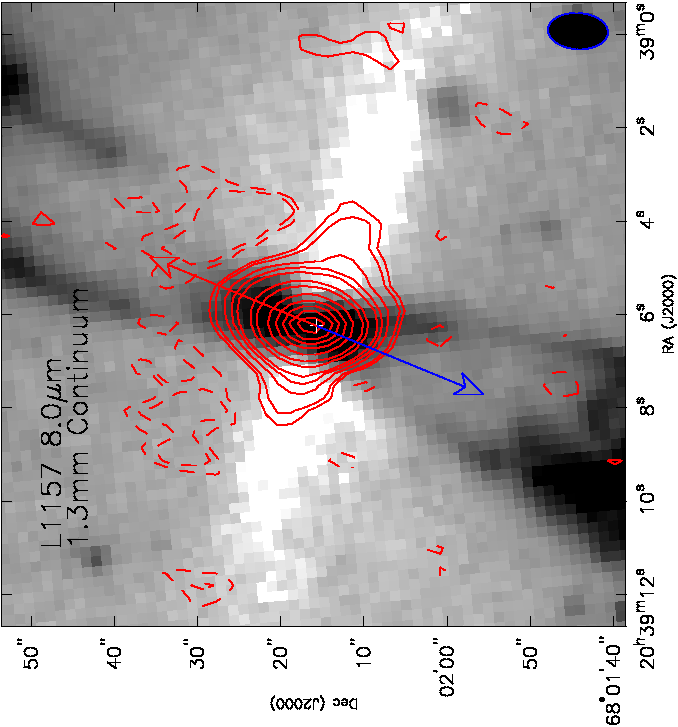}
\caption{
With higher sensitivity, SMA observations detect the extension of the
flattened envelope and the increased heating along the outflow cavity.
The blue and red arrows indicate the outflow directions of
the blue-shifted and red-shifted component, respectively.
The beam size is in the bottom-right corner.
The contours are $\pm2, \pm3, 6, 9, 12, 22, 30, 50\sigma$, where $\sigma=2.16$ mJy beam$^{-1}$.
}
\label{fig:sma}
\end{center}
\end{figure}

However, with enough sensitivity the filamentary structures can still be seen in dust emission.
%Figure \ref{fig:sma} is from \cite{stephens}, which observed dust polarization toward
%L1157 with SMA at $\lambda$~=~1.3~mm.  
Figure \ref{fig:sma} is the dust emission toward L1157 with the Submillimeter Array (SMA) at $\lambda$~=~1.3~mm (Tobin et al.\ 2012, in prep).
%Polarization detection requires
%longer integration time, increasing the sensitivity to the dust emission in general.
In this case, they detected the extension along the flattened envelope and even an extension
along the outflow (also see Stephens et al.\ 2012, in prep).  The extension along the outflow illustrates how the heating is facilitated
by lower density material (in this case in the outflow cavity). 
In other words, the heating in these sources are not uniform, which can lead to a distortion
in the structure suggested by only the dust continuum. 

\begin{figure*}
\begin{center}
\includegraphics[scale=1.00]{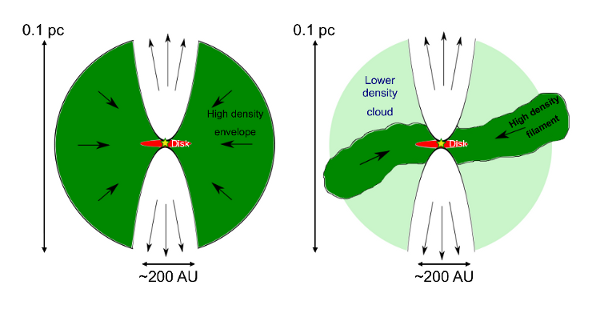}
\caption{Schematic view showing the changing view of star formation from spherical collapse (left panel) to filamentary collapse (right panel).  Note that the size scales are exaggerated to better
illustrate the structures.
}
\label{fig:class0}
\end{center}
\end{figure*}

Indeed, when comparing the observations of L673 and L1157 
with the traditional view of spherical star formation,
we need to change the cartoons of star formation.
Figure \ref{fig:class0} demonstrates our suggestion of moving from
spherical star formation structures to filamentary star formation structures
in Class 0 protostars to be more consistent with observations.  
The left panel presents the traditional model assuming sphericity that has 
impacted our theoretical understanding for decades.  
In this model, protostellar collapse is axisymmetric and spherical based on a singular isothermal sphere \citep{1977ApJ...214..488S}, 
With the inclusion of rotation \citep{1984ApJ...286..529T}, the density structure is slightly flattened and mostly remains spherical beyond the centrifugal radius.  
%and although rotation can change the inner density \citep{1984ApJ...286..529T},
%the source still remains mostly spherical.  
On the other hand, the right panel in Figure \ref{fig:class0} 
shows the axisymmetric and filamentary envelopes 
that are often seen in our Class 0 observations \citep[e.g.,][]{2010ApJ...712.1010T}.
The filamentary envelopes with higher density are forming inside the 
ambient cloud at lower density.

The change from spherical view of star formation to filamentary view certainly has
important
consequences, as several analysis techniques are based on the assumption of sphericity.  
For example, for a single beam measurement of the region the spherical assumption will significantly underestimate the mean density of the dense material. Alternatively, low sensitivity interferometric maps will concentrate on the peak and likely miss the large structure and thus the shape. 
In addition, blue-skewed spectra have been extensively observed with optically thick molecular lines in starless cores.  
The interpretation of the blue asymmetry, together with optically thin lines peaking in the absorption dip, has been spherical collapse.  
Moreover, the modeling of spectral energy distributions (SEDs) \citep[e.g.,][]{2003ApJ...591.1049W,2006ApJS..167..256R} extensively used in Class 0 and Class I sources is based on spherical/axisymmetric models. 
As non-spherical envelopes are more common, spherical models may not provide 
accurate descriptions of protostellar properties, so extra caution needs to be applied.

\subsection{Observationally-Driven Scenario for Filamentary Collapse}
\label{sec:evo}

\begin{figure}
\begin{center}
\includegraphics[scale=0.7]{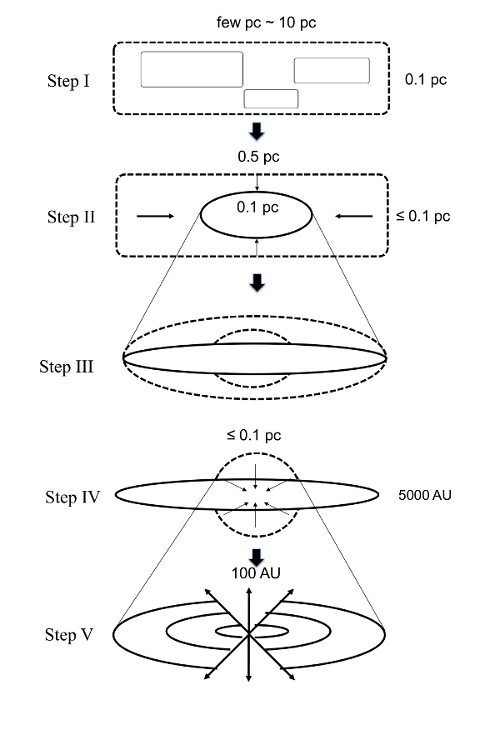}
\caption{
Illustration of filamentary collapse in five Steps.  In Step I, molecular clouds are formed
with filamentary shapes (few parsecs in length and 0.1 pc in width) that are prone to
fragmentation.  In Step II, the subsequently-fragmented filaments (few tenth of parsecs in
length) collapse and form prolate or oblate starless cores (0.1 pc in size).  In Step III,
embedded in the starless core is a filamentary structure of higher density that
arises from the flow along the the large-scale filament axis.
In Step IV, the starless core continues to infall into a centrally condensed envelope
of a Class 0 protostar
($\sim$ 5000 AU).  In Step V, the Class 0 source evolves to a Class I source with a protostellar
disk (few hundred AUs) and outflow (the arrows).
The orientation of the protostellar disk depends on the detailed kinematics of collapse and is not
necessarily along the filament as shown in this cartoon.
Note in each step the structures with lower density are indicated with dashed lines.
}
\label{fig:cartoon}
\end{center}
\end{figure}

Filamentary structures appear to be ubiquitous from large molecular clouds to small scale circumstellar envelopes.  
These filamentary structures are also observed to be tightly connected to the star formation process, as prestellar cores and young protostars are located within these filaments.  
From these observations, we propose an observationally derived scenario of filamentary collapse in star formation that is summarized in Figure \ref{fig:cartoon}.  
As shown in the cartoon, there are approximately five steps in our
observational-based picture of the filamentary collapse process.
Among the five steps, Step I, II, IV and V are from observations, and Step III is a prediction of high-density filamentary structures in starless cores, to connect Step II and IV.

In Step I, molecular clouds are formed as filaments with a 
few parsecs to 10 parsecs in length and a characteristic width of 0.1 pc \citep{1979ApJS...41...87S,1987ApJ...312L..45B,1999ApJ...510L..49J,2010A&A...518L.102A}.
These large-scale filaments are probably turbulent and prone to fragmentation, leading to subsequent velocity-coherent, higher density filaments (few tenths of parsecs in length) that are considered as the birthplaces of prestellar cores \citep{1997ApJ...480..681I,2002ApJ...578..914H,2011ApJ...740...88P}.
In Step II, the fragmented filaments collapse along the long and short axis 
while feeding material along the filament \citep[e.g.,][]{2011A&A...533A..34H},
enhancing the  mass in a location and forming a higher density oblate (or prolate) 
starless core as observed with single-dish observations 
\citep{2002ApJ...576..849C,2002ApJ...569..280J,2007MNRAS.379L..50T,2009MNRAS.399.1681T}.
The core formation may be related to the flows from large-scale motions 
along the larger filaments and is kinematically coupled with the parental cloud 
\citep{2011A&A...533A..34H}.  This would form a higher density filamentary
structure embedded inside of the starless core as seen in Step III.
This substructure is the kinematic descendant of the flow along the larger filament
and the origin of the filamentary envelopes seen in the Class 0 objects.
As the collapse continues in Step IV, 
material infalls along the smaller filament \citep{2012arXiv1201.2174T} 
and the oblate (or prolate) starless core continues to collapse into 
a centrally condensed envelope of a Class 0 protostar.
A Class 0 source is created ($\sim 5000$ AU in size), while the large scale filamentary structure ($\sim$ 1000 AU to 0.1 pc) remains behind containing an appreciable fraction of the total mass of the envelope plus source.
In Step V, the Class 0 source evolves to a Class I source with a protostellar disk and the larger structure dissipates.

%The filamentary protostellar envelopes observed in \citet{2010ApJ...712.1010T} corresponds to 
%step three.
%However, while the formation of these filamentary protostellar envelopes is still unclear.
%the structure must have begun to grow during the stage prior to protostellar phase, i.e., the prestellar phase.
%Single-dish observations of starless cores have established our knowledge toward prestellar cores in morphology, density structure, chemical abundances and kinematics
%\citep{1991ApJ...376..561M,1993ApJ...406..528G,1994MNRAS.268..276W,1998ApJ...504..900T,2002ApJ...569..815T}.
%In particular, studies with single-dish observations have demonstrated that majority of starless cores have flattened and elongated morphologies, either in prolate or oblate
%\citep{2002ApJ...569..280J,2007MNRAS.379L..50T}.
%These flattened structures beyond spherical symmetry suggest a transition from large-scale filaments in molecular clouds to small-scale filamentary protostellar envelopes.

Although this scenario fits together, there is one serious problem with our proposed
evolution of filaments in star formation:
no one has detected the substructure (e.g.\ filamentary structure) predicted in Step III
in starless cores to date.
Unfortunately, there is some difficulty in detecting these structures.  One could use molecular line
tracers such as N$_2$H$^+$ or NH$_3$, which often correspond to the 8 $\um$ absorption
\citep[e.g.,][]{2010ApJ...709..470C,tobin2011}.  However in starless cores, 
N$_{2}$H$^{+}$ could still have chemical effects such as depletion \citep[e.g.,][]{ted2002}, 
although several studies showed \citet{2002ApJ...569..815T} less depletion for N$_{2}$H$^{+}$ than other molecules.
Since the depletion usually ocurrs at the center of the core, 
the filament could appear fragmented in the map depending on the size of the central depletion.  
In addition, the molecular distribution could
originate from chemistry and not well-trace the dense material.  
Thus, to confirm detection of substructure, we must rely on dust continuum emission.
Dust emission at millimeter wavelengths presumably is more appropriate than 8 $\um$ extinction because of the low optical depth. 
8 $\um$ extinction shows detection in the outer regions only if the background signal to noise is high enough.

In order to resolve the structures, we must have resolution of $\sim$5 arcsec, which implies interferometers.
For example, \citet{2010ApJ...718..306S} performed dust continuum observations at 3 mm toward 11 starless cores in Perseus 
with CARMA.
Although two sources were detected, they were later reclassified as protostellar objects 
\citep{2010ApJ...722L..33E,2012ApJ...745...18S}, implying only non-detections of sub-structure of starless cores,
contrary to our suggested evolutionary sequence.
Our explanation is that the sub-structure was not detected due to a lack of sensitivity.
In the following section, we investigate that possibility and place constraints on the 
underlying filamentary structure based on \citet{2010ApJ...718..306S} results.

\section{Synthetic Observations}
\label{sec:obs}

To examine the likelihood of our proposed structures in starless cores,
we make synthetic observations with CARMA, directly
comparing to the observations of \citet{2010ApJ...718..306S}, and ALMA, using
the flattened envelope around L1157 as a model.
The result will show that the expected structures are below CARMA D array's detection threshold at 3mm, but they
should be detectable with CARMA D and E array observations at 3mm, CARMA E array at 1mm, and ALMA 1mm observations.
This implies that there is not yet a disagreement between our observational-based proposed evolutionary scheme in Step III for low-mass star formation and current observations, 
and an exciting observational future is suggested. 
%This implies that there is no problem with our observational-based proposed evolutionary scheme for low-mass star
%formation and an exciting observational future.

\subsection{CARMA observations}

We simulate CARMA imaging with parameters used by \citet{2010ApJ...718..306S}:
heterogeneous array (six 10-meter antennas and nine 6-meter antennas)
imaging with the CARMA-D array configuration at 3 mm continuum.
We use the Miriad tasks \textit{uvgen}, \textit{demos} and \textit{uvmodel}, 
based on \citet{2010SPIE.7733E..37W} without the 3.5-meter telescopes.  
To find the detection limit of such a structure, we also simulate CARMA D+E array observations at 3 mm 
and CARMA E array at 1 mm.  
Baselines range from 3$k \lambda$ to 38$k \lambda$ for the D array at 3 mm, 
2$k \lambda$ to 19$k \lambda$ for the E array at 3 mm,
and 5$k \lambda$ to 47$k \lambda$ for the E array at 1 mm. 
The observing rest frequency is centered at 90 GHz for 3 mm observations and 230 GHz for 1 mm observations with a total bandwidth of 4 GHz for continuum observations.
The total observing time on the target is 6 hours for each synthetic observation (for the one with CARMA D+E array, the observing time is three hours for the D array and three hours for the E array).  
In the analysis (Sect \ref{sec:L1157}), we perform a small mosaic (standard seven-pointings) around the source to capture all the extended structure.
Table \ref{tbl:obs} summarizes the synthesized beam sizes and noise levels for the simulated CARMA observations.  

\begin{deluxetable*}{lccc}
\tablecaption{Synthesized beam sizes and noise levels for the synthetic observations}
\tabletypesize{\small}
\tablewidth{0pt}
\tablecolumns{4}
\tablehead{
\colhead{} & \colhead{CARMA D array} & \colhead{CARMA D+E array} & \colhead{CARMA E array} \\
\colhead{} & \colhead{3 mm} & \colhead{3 mm} & \colhead{1 mm}
}
\startdata
L1157 modeling (Sect \ref{sec:L1157}) & $5.14\arcsec \times 4.76\arcsec$ & $7.92\arcsec \times 7.23\arcsec$ & $4.21\arcsec \times 3.65\arcsec$ \\
  & $0.3$ mJy beam$^{-1}$ & $0.15$ mJy beam$^{-1}$ & $0.35$ mJy beam$^{-1}$ \\
\enddata
\label{tbl:obs}
\end{deluxetable*}

\subsection{ALMA observations}
We used the task \textit{sim\_observe} and \textit{sim\_analyze} in the package \textit{casapy} to perform the simulated observations with ALMA.  
The angular resolution is requested to be 1.2$\arcsec$ in the simulation to observe detailed structures, and a small mosaic is applied to capture all possible structure.
The observing time for each mosaic pointing is 100 s, and the total observing time is 2 hours. 
The observing frequency is centered at 90 GHz with the bandwidth of 8 GHz for continuum observations. 
Thermal noise is added with a typical precipitable water vapor of 2.8 mm. 
The clean threshold is set to 1.5 times the noise rms, and the pixel size is set to 0.12 arcsecs. 

\subsection{Modeling and Results}
\label{sec:analyze} 
\label{sec:L1157}

\begin{figure*}
\begin{center}
\includegraphics[scale=0.5,angle=270]{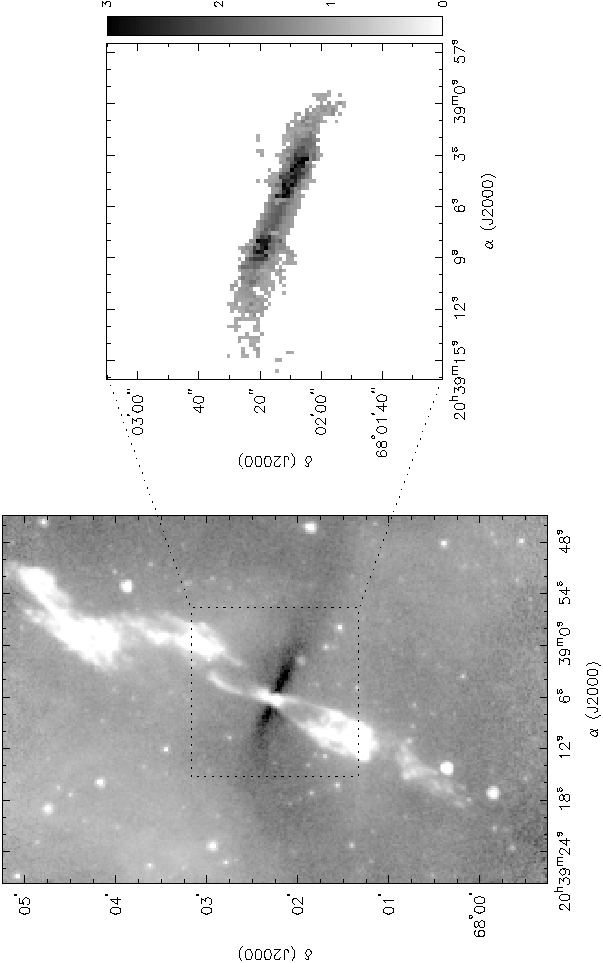}
\caption{(a): the left panel.  It is the extinction map of L1157 from the Spitzer 8 $\mu m$ observation.  A flattened envelope is seen in absorption against the background emission around the central Class 0 object with the outflow nearly perpendicular to it.  (b): the right panel.  The emission from the outflow and the scattered light from the central object are removed.  The central part is filled with the average of the envelopes on the two sides. The color scale shows optical depth.}
\label{fig:spitzer}
\end{center}
\end{figure*}

As decribed in Step III and IV in Figure \ref{fig:cartoon}, the flattened protostellar envelopes around Class 0 sources are speculated to be highly connected to the filamentary structure at the previous stage, the prestellar phase.  
Therefore, the envelopes around Class 0 sources best describe the morphology of the high-density filamentary structures in Step III.  
To simulate the structure in Step III, 
we modify the flattened envelope in the Class 0 source L1157 
by the physical conditions expected at the prestellar stage, 
in order to examine if CARMA and ALMA are able to detect filamentary structure at the prestellar stage. 
Figure \ref{fig:spitzer}(a) shows the extinction map of L1157 from the Spitzer 8$\micron$ observation \citep{2007ApJ...670L.131L}.  
We chose L1157 to model as it has an obvious filamentary envelope structure seen in the 8$\micron$ absorption against the background emission, and the symmetric structure can be approximated with a radial density power-law  \citep{2007ApJ...670L.131L}.  
The spatial scale of the filamentary envelope is 0.1 pc, too large for a circumstellar disk or a pseudo disk \citep[e.g.,][]{1993ApJ...417..243G}. 
The L1157 dark cloud is located $\sim$ 250 parsecs\footnotemark\ away with an edge-on view concealing the Class 0 source embedded in the
flattened envelope nearly perpendicular to a large powerful outflow from the north to the south. 
The distance to L1157 is approximately the same as Perseus (also at $\sim$ 250 pc) and thus remains an excellent proxy for Perseus when comparing with \citet{2010ApJ...718..306S}.  
If the proposed scenario of filamentary collapse is correct, the flattened envelope is expected to be related to the filamentary structures on larger scales, and thus this source is suitable for modeling the transient phase in the prestellar stages with appropriate physical conditions.

\footnotetext{The L1157 cloud is estimated to have a similar galactic latitude as the absorbing clouds with 200 pc and 300 pc in Cepheus \citep{1998ApJS..115...59K} and therefore we adopted a distance of 250 pc in this paper. 
In comparison, \citet{2009ApJS..185..198K} adopted a distance of 325 pc for the region around L1157.  
}
 
To better concentrate on the filamentary envelope, we removed the emission from the outflow and the scattered light from the central object.  
We then filled the inner regions with the averaged value from the envelopes on the two sides, as shown in Figure \ref{fig:spitzer}(b). 
The total mass calculated from the extinction increased by about 10\% by filling the inner region with this method.
The extinction map was compared toward background stars measured in the near-IR to an optical-depth image generated with or without the zodiacal correction \citep{2010ApJ...712.1010T}.  

We next generated the brightness map at millimeter wavelengths (at 3 mm in our model) by assuming that L1157 is optically thin at 3 mm.  
%This assumption is justified by the long wavelength.  
%reasonable since the optical depth at 8 $\um m$ is $\sim$ 2 in average from the extinction map and the optical depth at 3 mm decreases by a factor of 6500 according to the relation $\tau_{3mm} / \tau_{8 \mu m} = \kappa_{3mm} / \kappa_{8 \mu m}$ (see the formula and numbers for $\kappa_{3mm}$ and $\kappa_{8 \mu m}$ below).       
We first calculated the mass contained in each pixel from the extinction map,

\[
M = d\Omega \times D^{2} \times \left( 1.496 \times 10^{13} \times \frac{{\rm cm}}{{\rm AU}} \right) ^{2} \times \frac{\tau}{\kappa_{8\mu m}},
\]
where $d\Omega$ is the pixel solid angle $(1.2\arcsec)^{2}$, D is the distance in parsecs (250 pc for L1157) and $\kappa_{8\mu m}$ is the dust plus gas opacity at $8\mu m$.
The 3\,mm flux in each pixel is then determined from the mass, opacity, and temperature, 

\[
F = \frac{M \times B_{\nu}(T) \times \kappa_{3mm}}{D^{2}}, 
\]
where $B_{\nu}(T)$ is the Planck function and $\kappa_{3mm}$ is the dust plus gas opacity at 3mm.

\begin{figure*}
\begin{center}
\includegraphics[scale=0.7,angle=270]{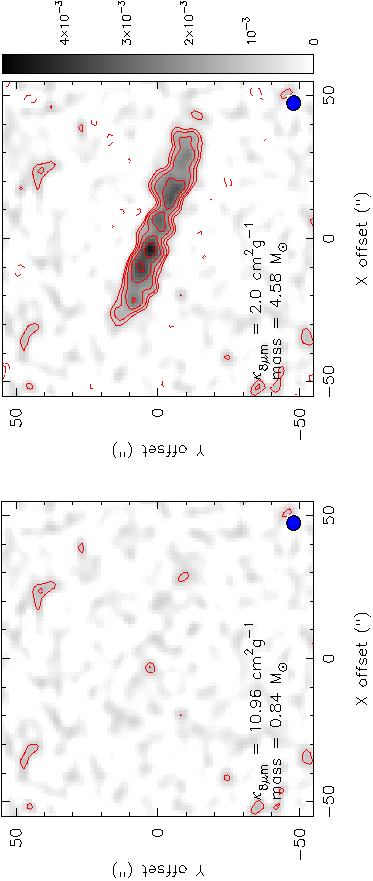}
\caption{Simulated observations with CARMA D array at 3 mm for six hours with different values of $\kappa_{8\mu m}$.  The center is positioned at $20^{h}39^{m}05.2^{s}$ (RA) and $68\degr02\arcmin15.3\arcsec$ (Dec) (J2000).  The synthesized beam size is 5.14\arcsec by 4.76 \arcsec shown on the bottom-right corner.  The noise level $\sigma$ is 0.3 mJy/beam, and the contours are $\pm3, \pm4.2, \pm6, \pm8.5, \pm12, \pm17, \pm24, \pm34 \times \sigma$ (in step of $\sqrt{2} \sigma$).  The color scale shows flux in Jy/beam.  With the reasonable value for $\kappa_{8\mu m}$ (10.96 cm$^{2}$ g$^{-1}$) in the left panel, no structures are detected.  For the structures to be clearly detected (the right panel), $\kappa_{8\mu m}$ needs to be an almost impossibly small value (2.0 cm$^{2}$ g$^{-1}$). }
\label{fig:L1157}
\end{center}
\end{figure*}

We used 0.00169 cm$^{2}$ g$^{-1}$ for $\kappa_{3mm}$ by assuming 100 for the gas to dust ratio \citep[e.g.,][]{2010ApJ...718..306S}.  
The temperature was assumed to be a constant at 10 K for starless cores \citep[e.g.,][]{2009ApJ...691.1754S}.  After obtaining the brightness map at 3 mm, we simulated the CARMA D array observations with our model.
Since the opacity in the infrared is poorly constrained, 
we generated models with varying values for $\kappa_{8\mu m}$, 
which also modifies the derived mass in the flattened envelope structure.
The equation above indicates that the observed millimeter brightness decreases with increasing $\kappa_{8\mu m}$, since less mass is required to produce the IR extinction.
%To see how the observed brightness changes, under these assumptions, we show the simulated results with CARMA D array for the six cases with $\kappa_{8\mu m} =$ (2.0, 4.0, 5.9, 7.0, 8.0, 10.96) in Figure \ref{fig:L1157}.   
%Each value of $\kappa$ corresponds to a value for the total mass (4.58, 2.29, 1.55, 1.31, 1.15, 0.84) $M_{\sun}$.  
%As the results show, the millimeter brightness decreases with increasing $\kappa_{8\mu m}$, since less mass is required to produce the IR extinction. i
%The limit for detection is $\kappa_{8\mu m}$ $\sim$ 7.0 cm$^{2}$ g$^{-1}$, and there is almost no detection for $\kappa$ larger than 7.0 cm$^{2}$ g$^{-1}$.  
%As the reasonable value for $\kappa_{8\mu m}$ toward L1157 should be $\sim$ 10.96 cm$^{2}$ g$^{-1}$ \citep{2010ApJ...712.1010T,2009ApJ...696..484B}, these results suggest that with a reasonable dust opacity for starless cores, the CARMA-D array is not able to detect such filamentary structures at the prestellar stage. 
In the left panel of Figure \ref{fig:L1157}, we show that with an expected value of $\sim$ 10.96 cm$^{2}$ g$^{-1}$ for $\kappa_{8\mu m}$ \citep{2010ApJ...712.1010T,2009ApJ...696..484B} toward L1157, the CARMA-D array is not able to detect the filamentary structures at the prestellar stage.
For the structures to be clearly detcted (the right panel of Figure \ref{fig:L1157}), the dust opacity at 8 $\um$ would have to be an unphysically small value.

\begin{figure}
\begin{center}
\includegraphics[scale=0.35,angle=270]{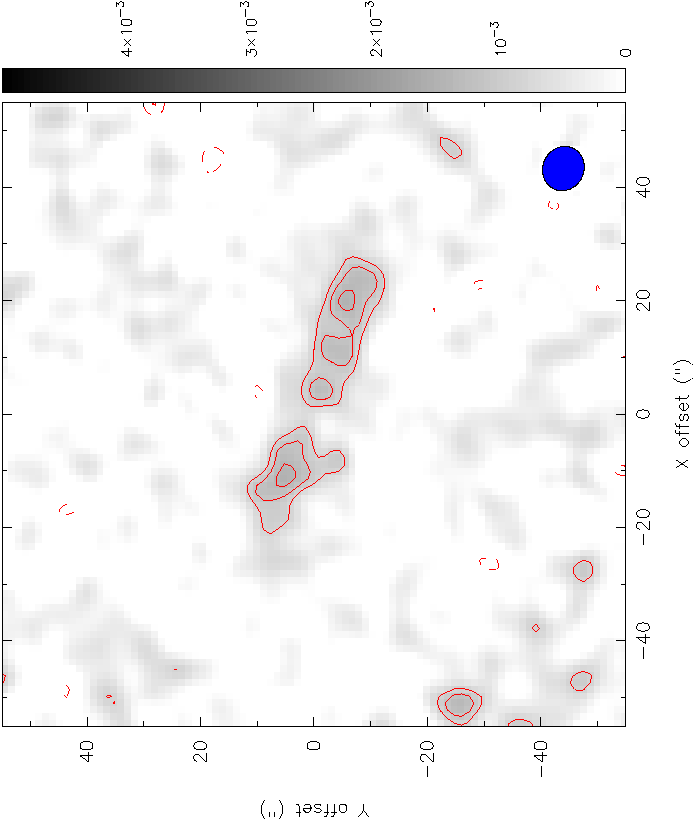}
\caption{Simulated observations with CARMA D+E array at 3 mm for six hours in total (three hours for the D array and three hours for the E array).  The value for $\kappa_{8\mu m}$ is 10.96 cm$^{2}$ g$^{-1}$.  The center is positioned at $20^{h}39^{m}05.2^{s}$ (RA) and $68\degr02\arcmin15.3\arcsec$ (Dec) (J2000).  The synthesized beam size is 7.92\arcsec by 7.23\arcsec shown on the bottom-right corner.  The noise level $\sigma$ is 0.28 mJy/beam, and the contours are $\pm3, \pm4, \pm5, \pm6, \pm7, \pm8, \pm9, \pm10 \times \sigma$.  The color scale shows flux in Jy/beam.}
\label{fig:L1157_DE}
\end{center}
\end{figure}

\begin{figure}
\begin{center}
\includegraphics[scale=0.35,angle=270]{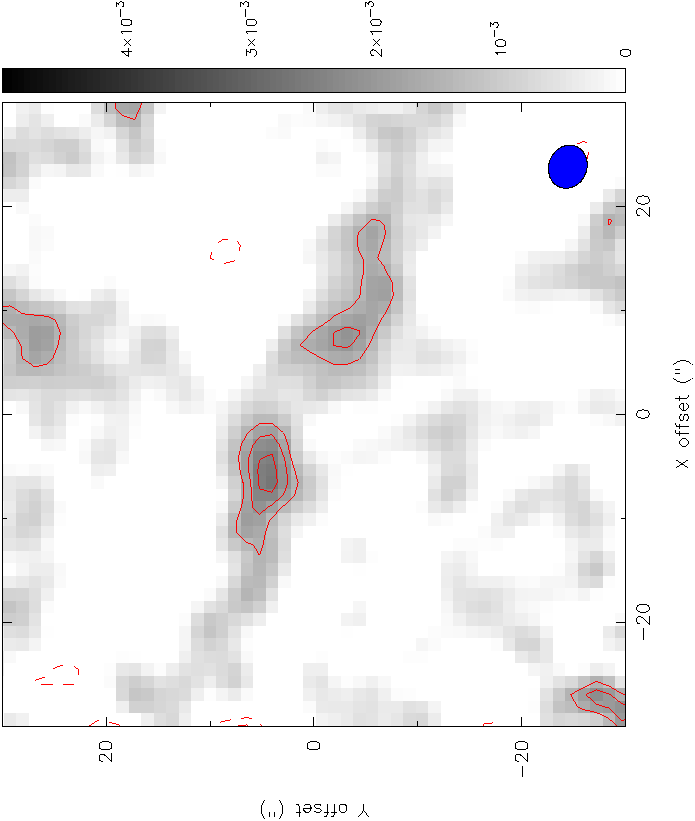}
\caption{Simulated observations with CARMA E array at 1 mm for six hours in total.  The value for $\kappa_{8\mu m}$ is 10.96 cm$^{2}$ g$^{-1}$.  The center is positioned at $20^{h}39^{m}05.2^{s}$ (RA) and $68\degr02\arcmin15.3\arcsec$ (Dec) (J2000).  The synthesized beam size is 4.21\arcsec by 3.65\arcsec shown on the bottom-right corner.  The noise level $\sigma$ is 0.49 mJy/beam, and the contours are $\pm3, \pm4, \pm5, \pm6, \pm7, \pm8, \pm9, \pm10 \times \sigma$.  The color scale shows flux in Jy/beam.}
\label{fig:L1157_1mm}
\end{center}
\end{figure}

To fully explore CARMA's capability, we performed the synthetic observation with CARMA D+E array at 3 mm since the E array is more compact and sensitive to emission at larger scales than the D array.  
The total observing time is six hours (three hours with the D array and three hours with the E array).  
The value for $\kappa_{8\mu m}$ used is 10.96 cm$^{2}$ g$^{-1}$ to compare with the result from the D array only, since it produces the weakest emission (contains the least mass).
As shown in Figure \ref{fig:L1157_DE}, the filamentary structure is detected with a similar noise level (0.28 mJy/beam) as the D array (0.3 mJy/beam), 
although only the structures with stronger emissions close to the center could be seen and the structures are not in detail.  
The detection suggests that the non-detection with the D array is due to a combination of spatial resolution and sensitivity.  
Furthermore, we shift the observation from 3 mm to 1 mm assuming that the dust opacity is 0.9 cm$^{2}$ g$^{-1}$ at 1 mm \citep{1994A&A...291..943O} and $\kappa_{8\mu m}$ is still 10.96 cm$^{2}$ g$^{-1}$. 
Again, CARMA E array is able to show detection on the structure as shown in Figure \ref{fig:L1157_1mm}, as the brightness increases toward the short wavelengths.

\begin{figure}
\begin{center}
\includegraphics[scale=0.33,angle=270]{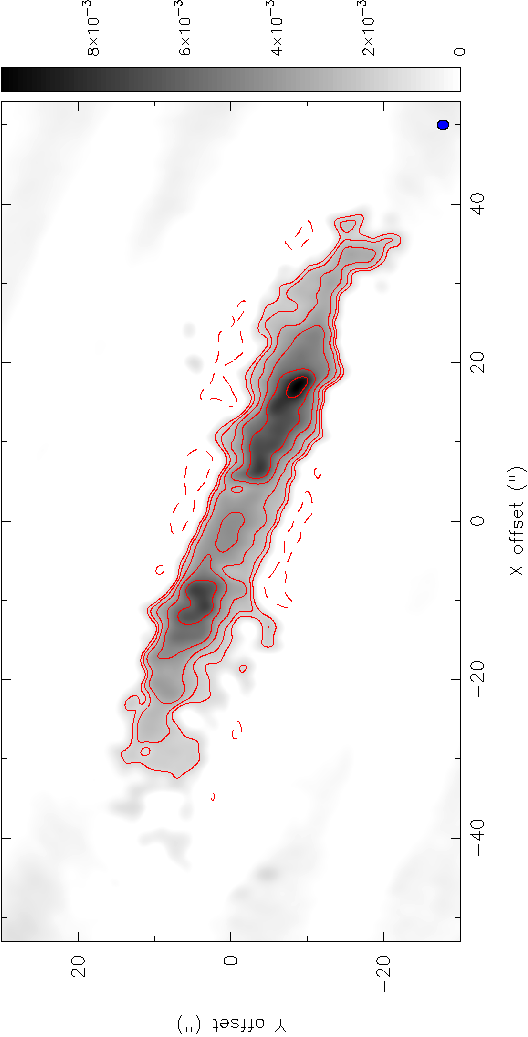}
\caption{Simulated observations with ALMA for $\kappa_{8\mu m} = 10.96$ cm$^{2}$ g$^{-1}$.  The noise level $\sigma$ is 0.15 mJy beam$^{-1}$.  The contours indicate $\pm10, \pm15, \pm20, \pm25, \pm30, \pm40, \pm50, \pm60 \times \sigma$ .  The color scale shows flux in Jy/beam.  The synthesized beam size is $1.2\arcsec$ shown on the bottom-right corner.}
\label{fig:L1157_alma}
\end{center}
\end{figure}

We further ran the synthetic observations with ALMA also for the case of $\kappa_{8\mu m} = 10.96$ cm$^{2}$ g$^{-1}$.
Figure \ref{fig:L1157_alma} shows the result from the synthetic observation (the contours are plotted by percentages of the peak flux instead of the noise levels due to the artificial effects from resolving-out large structures). 
As can be clearly seen, ALMA is able to detect most of the structure in the flattened envelope.  
Comparing the results from CARMA and ALMA for $\kappa_{8\mu m} = 10.96$ cm$^{2}$ g$^{-1}$, ALMA's unprecedented sensitivity greatly improves the appearance of filamentary structures and should provide a powerful tool for uncovering any hidden filamentary profiles at the starless/prestellar stage.
Located in the southern hemisphere, ALMA is not able to look at L1157; 
these results are, however, indicative of the ALMA observations with other starless cores.

\section{Conclusion}

In this paper, we posit an observationally derived scenario for
filament-driven star formation that incorporates the evolution of
star-forming cores with filaments into filamentary envelopes from large to small scales.
Molecular clouds are formed as filaments (few parsecs to 10 parsecs)
and then fragment to smaller filaments (few tenths of parsecs),
which eventually collapse to form triaxial starless cores.  As
collapse continues the material infalls along the filament into a
centrally condensed filamentary envelope of a spherical Class 0 source, which
keeps evolving to a Class I source with a protoplanetary disk.

If such a scenario is correct, the filamentary structures at the prestellar stage should exist.  
The only reason that they have not yet been detected is sensitivity to large-scale emission in the surveys.
They are possible to detect with CARMA D+E array at 3 mm due to more appropriate resolution and CARMA E array at 1 mm 
as a result from the higher brightness at 1 mm; however, ALMA is even more capable of clearly detecting detailed structures of the filamentary envelopes.  
In fact, the very high sensitivity of ALMA will allow for much shorter integrations (less than two hours for L1157-like prestellar cores) and thus we will be able to conduct quick and efficient surveys of the geometry of the envelopes around starless cores.  The scenario proposed scenario can be
immediately tested by observations with the current instruments.

\section{Acknowledgements}

We thank the anonymous referee for valuable comments to improve this paper.
We acknowledge support from the Laboratory for Astronomical Imaging
at the University of Illinois.
Support for CARMA construction was derived from the states
of Illinois, California and Maryland, the Gordon and Betty Moore
Foundation, the Eileen and Kenneth Norris Foundation, the Caltech
Associates and the National Science Foundation. Ongoing CARMA
development and operations are supported by the National Science
Foundation, and by the CARMA partner universities.
D.J.\ acknowledges support from an NSERC Discovery Grant. 
J.\ Tobin acknowledges support provided by NASA through Hubble Fellowship
grant \#HST-HF-51300.01-A awarded by the Space Telescope Science Institute, 
which is operated by the Association of Universities for Research in Astronomy,
Inc., for NASA, under contract NAS 5-26555.
J.\ Tobin also acknowledges support from the National Radio Astronomical Observatory.
The National Radio Astronomy Observatory is a facility of the National Science
 Foundation operated under cooperative agreement by Associated Universities, Inc.

\bibliographystyle{apj}
\bibliography{paper.bib}
%\clearpage
%\input{fig}
%\clearpage

\end{document}